\documentclass[prb,twocolumn]{revtex4-1}

\usepackage{amssymb}
\usepackage{amsmath}
\usepackage{graphicx}

\begin{document}

\title{Magnetostatic interactions between magnetic nanotubes}
\author{J. Escrig, S. Allende, D. Altbir}
\author{M. Bahiana}
\affiliation{Departamento de F\'{\i}sica, Universidad de Santiago de Chile, USACH, Av.
Ecuador 3493, Santiago, Chile}
\affiliation{Instituto de F\'{\i}sica, Universidade Federal do Rio de Janeiro, CP 68528,
21941-972 Rio de Janeiro, Brazil}

\begin{abstract}
The investigation of interactions between magnetic nanotubes is complex and
often involves substantial simplifications. In this letter an analytical
expression for the magnetostatic interaction, taking into account the
geometry of the tubes, has been obtained. This expression allows for the
definition of a critical vertical separation for relative magnetization
between nanotubes and can be used for tailoring barcode-type nanostructures
with prospective applications such as biological separation and transport.
\end{abstract}

\maketitle

Since the discovery of carbon nanotubes by Iijima in 1991, \cite{Iijima91}
intense attention has been paid to tubular nanostructures. Because of their
geometry, nanotubes offer prospective applications in catalysis, \cite%
{SCG+04, MLT+02} sensors, \cite{KNS02} and biological separation and
transport. \cite{MLT+02, LMT+02, KHC+04, SRH+05} In particular, magnetic
nanotubes are the focus of growing interest due to the existence of
techniques that lead to the production of highly ordered arrays. \cite%
{DKG+07, BJK+07, EBJ+08} These particles offer an additional degree of
freedom as compared to nanowires; not only can the length, $L$, and external
radius, $R$, be varied, but also the internal radius, $a$. In this way,
nanotubes may be suitable for applications in biotechnology, where magnetic
nanostructures with low density, which can float in solutions, become much
more useful for in vivo applications. \cite{Eisenstein05} These tiny
magnetic tubes could provide an unconventional solution for several research
problems, and a useful vehicle for imaging and drug delivery applications. 
\cite{SRH+05, XCV+08}

In such systems changes in thickness are expected to strongly affect the
mechanism of magnetization reversal \cite{LAE+07} and, thereby, the overall
magnetic behavior. Also interactions play a fundamental role, modifying the
magnetic behavior of the particles. Clearly, for the development of magnetic
devices based on those arrays, knowledge of the magnetostatic interaction
between the tubes is of fundamental importance. But as usual the effects of
interparticle interactions are complicated by the fact that the dipolar
field felt by each element depends upon the magnetization state of all the
elements in the array. In a previous work by Lee \textit{et. al.}, \cite%
{LSN+05} multisegmented metallic nanotubes with a bimetallic stacking
configuration along the nanotube axes were prepared and investigated. These
particles exhibit different magnetic behaviors, which encourage a study
about the magnetostatic interactions between the stacking. Due to the very
narrow hysteresis loops that are obtained, the influence of the interactions
is not easily identifiable from magnetization curves, and then a theoretical
study can shine light on this problem.

The purpose of this letter is to develop an analytical model for the full
long-range magnetostatic interaction between two nanotubes exploring the
possibility of varying the magnetic coupling as a function of the tubes
position. The geometry of the tubes is characterized by their external and
internal radii, $R$ and $a$, respectively, and length $L$. It is convenient
to define the ratio $\beta \equiv a/R$, so that $\beta =0$ represents a
solid cylinder (wire) and $\beta $ close to $1$ corresponds to a tube with
very thin walls. The separation between the tubes is written in terms of the
inter-axial distance, $d$, and the vertical separation, $s$, as depicted in
Fig. 1. Our model goes beyond the dipole-dipole approximation and lead us to
obtain an analytical expression for the interaction in which the lengths and
radii of the tubes are taken into account. We focus on the stability of
parallel and antiparallel magnetization alignment in pairs of interacting
tubes, as a function of the distance between them, in order to gain insight
on the understanding of the role of interactions on barcode-type nanotubes. 
\cite{NFR+01, LSN+05}

\begin{figure}[h]
\begin{center}
\includegraphics[width=6cm]{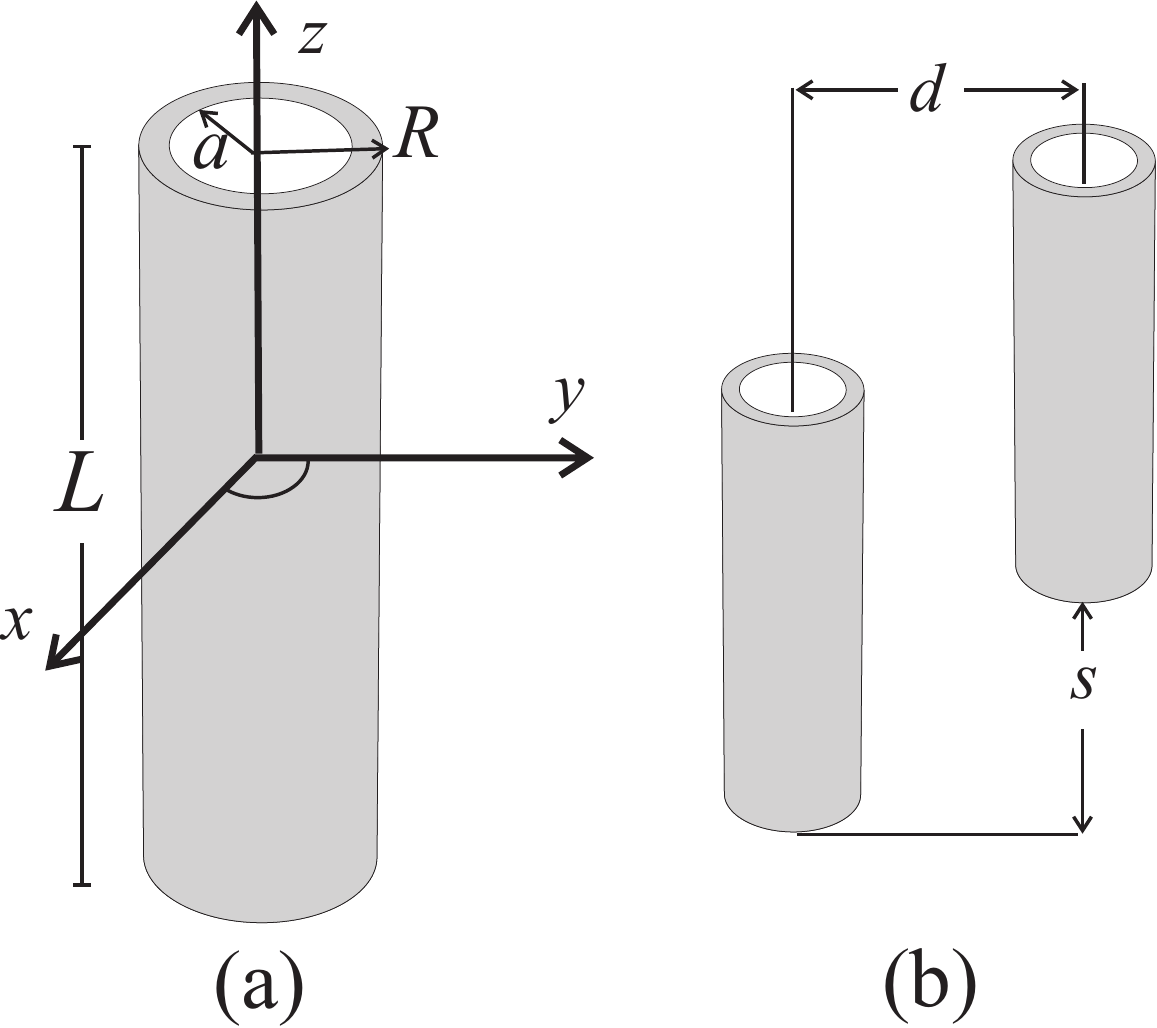}
\end{center}
\caption{(a) Geometric parameters used for the individual tube description.
(b) Relative position of interacting tubes: $d$ is the inter-axial distance
and $s$ the vertical separation.}
\end{figure}

We adopt a simplified description of the system, in which the discrete
distribution of magnetic moments is replaced with a continuous one
characterized by a slowly varying magnetization $\mathbf{M}\left( \mathbf{r}%
\right) $. The total magnetization can be written as $\mathbf{M}\left( 
\mathbf{r}\right) =\mathbf{M}_{1}\left( \mathbf{r}\right) +\mathbf{M}%
_{2}\left( \mathbf{r}\right) $, where $\mathbf{M}_{1}\left( \mathbf{r}%
\right) $ and $\mathbf{M}_{2}\left( \mathbf{r}\right) $ are the
magnetization of tubes $1$ and $2$, respectively. In this case, the
magnetostatic potential $U\left( \mathbf{r}\right) $ splits up into two
components, $U_{1}\left( \mathbf{r}\right) $ and $U_{2}\left( \mathbf{r}%
\right) $, associated with the magnetization of each individual tube. Then,
the magnetostatic energy of two interacting magnetic tubes may be written in
terms of their magnetizations and the fields generated by each one. The
general expression, after using the reciprocity theorem, is $E_{%
\mbox{\tiny
d}}= E_{\mbox{\tiny self}}^{1}+E_{\mbox{\tiny self}}^{2}+E_{\mbox{\tiny int}%
} $. The $E_{\mbox{\tiny self}}^{i}=\frac{\mu _{0}}{2}\int \mathbf{M}%
_{i}\left( \mathbf{r}\right) \cdot \mathbf{\nabla }U_{i}\left( \mathbf{r}%
\right)  dV_{i}$ terms correspond to the self-energy of the $i$-th
tube, and $E_{\mbox{\tiny int}}=\mu _{0}\int \mathbf{M}_{2}\left( \mathbf{r}%
\right) \cdot \mathbf{\nabla }U_{1}\left( \mathbf{r}\right) dV_{2}$
is the interaction energy between tubes, which is the focus of this letter.

In order to proceed, we first need to calculate the magnetostatic potential $%
U\left( \mathbf{r}\right) $ of a single tube. However, it is necessary to
specify the functional form of the magnetization for each nanotube. Due to
their geometry and in order to reduce the stray field, tubes present what is
called a flower configuration. \cite{HK02} In this case, the magnetization
is mostly homogeneous, and spreads outwards near the ends. This
non-homogeneity produces a decrease of the interaction energy felt by one
tube due to the other. For $L\gg R$ this decrease is small and can be
neglected to simplify the calculations. Thus, we consider tubes with an
axial magnetization defined by $\mathbf{M}_{i}\left( \mathbf{r}\right)
=M_{0}\sigma _{i}\mathbf{\hat{z}}$, where $M_{0}$ is the saturation
magnetization of each nanotube, $\mathbf{\hat{z}}$ is the unit vector
parallel to the axis of the nanotube and $\sigma _{i}$ takes the values $\pm
1$, allowing the magnetization of tube $i$ to point up $\left( \sigma
_{i}=+1\right) $ or down $\left( \sigma _{i}=-1\right) $ along $\mathbf{\hat{%
z}}$. The magnetostatic potential produced by the tube $1$ with
magnetization $\mathbf{M}_{1}(\mathbf{r})$ is given by volume and surface
contributions and can be written as 
\begin{equation}
U_{1}(\mathbf{r})=\frac{1}{4\pi }\left[ -\int_{V_{1}}\frac{\mathbf{\nabla }%
\cdot \mathbf{M}_{1}\left( \mathbf{r}^{\prime }\right) }{\left\vert \mathbf{r%
}-\mathbf{r}^{\prime }\right\vert }\,dV^{\prime }+\int_{S_{1}}\frac{%
\mathbf{\hat{n}}^{\prime }\cdot \mathbf{M}_{1}\left( \mathbf{r}^{\prime
}\right) }{\left\vert \mathbf{r}-\mathbf{r}^{\prime }\right\vert }\,dS^{\prime }\right] \,.  \label{u1}
\end{equation}%
Note that the first term on the right-hand side of Eq.~(\ref{u1}) vanishes
because the magnetization field is constant. Furthermore, the surface
integral in Eq.~(\ref{u1}) has contributions only from the upper and lower
ends of the tube, located at $z=L/2$ and $z=-L/2$, respectively. Due to the
symmetry of the problem, we have calculated the integral in (\ref{u1}) using
a kernel in cylindrical coordinates. \cite{Jackson} After some manipulation,
the integral expression for the scalar potential can be written as 
\begin{multline}
U_{1}\left( r,z\right) =\sigma _{1}\frac{M_{0}}{2}\int_{0}^{\infty }\frac{dk}{k}J_{0}\left( kr\right)  \\
\left[ RJ_{1}\left( kR\right) -aJ_{1}\left( ka\right) \right] \left(
e^{-k\left\vert \frac{L}{2}-z\right\vert }-e^{-k\left\vert -\frac{L}{2}%
-z\right\vert }\right) \,,
\end{multline}%
where $J_{n}\left( x\right) $ is the $n$-th order Bessel function of first
kind. Figure 2 shows the surface plot of $U_{1}$ for a thin-walled tube ($%
\beta =0.8$) and a wire ($\beta =0$). From this figure it is clear that one
should expect distinct behaviors from these two nanoelements in the region
near the ends.

\begin{figure}[h]
\begin{center}
\includegraphics[width=8cm]{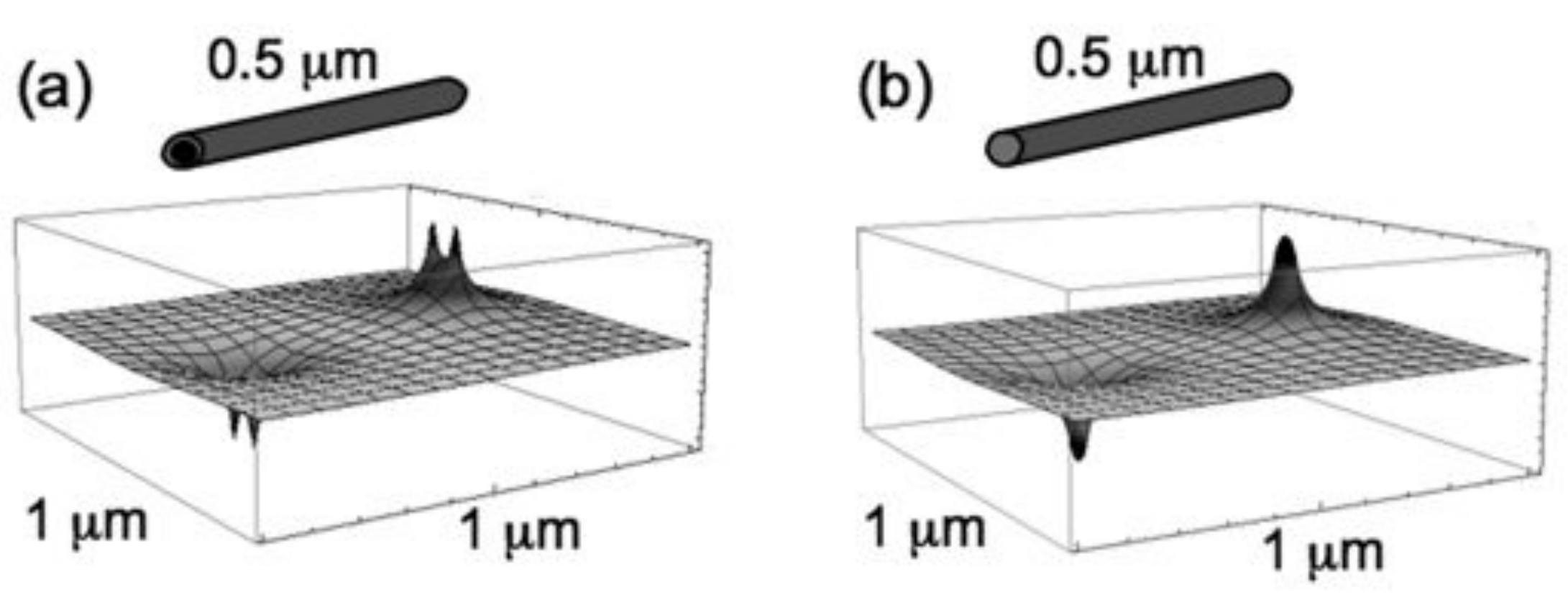}
\end{center}
\caption{Magnetostatic potential $U\left( \mathbf{r}\right) $ according to
equation (2) produced by a tube with $L=500$ nm, $R=50$ nm, and (a) $\protect%
\beta =0.8$ (thin-walled tube) and (b) $\protect\beta =0$ (wire). The color
scale is chosen such that higher absolute values of $U$ are represented by
darker shades.}
\end{figure}

Now it is possible to calculate the magnetostatic interaction energy between
two identical nanotubes using the magnetostatic field experienced by one of
the tubes due to the other. The final result reads 
\begin{multline}
\tilde{E}_{\mbox{\tiny int}}=-\frac{\sigma _{1}\sigma _{2}}{\left( 1-\beta
^{2}\right) }\int_{0}^{\infty }\frac{dq}{q^{2}}J_{0}\left( q\frac{d%
}{L}\right) e^{-q\left( 1+\frac{s}{L}\right) } \\
\left[ J_{1}\left( q\frac{R}{L}\right) -\beta J_{1}\left( q\frac{R}{L}\beta
\right) \right] ^{2}\left\{ 
\begin{array}{c}
\left( 1-e^{q}\right) ^{2}\qquad s\geq L \\ 
\left( 1-2e^{q}+e^{2q\frac{s}{L}}\right) \ s\leq L%
\end{array}%
\right. 
\end{multline}%
Here $\tilde{E}_{\mbox{\tiny int}}$ is the interaction energy in units of $%
\mu _{0}M_{0}^{2}V$, i.e. $\tilde{E}_{\mbox{\tiny int}}\equiv E_{%
\mbox{\tiny
int}}/\mu _{0}M_{0}^{2}V$, where $V=\pi R^{2}L\left( 1-\beta ^{2}\right) $
is the volume of a tube. Equation (3) has been previously obtained for
nanowires ($\beta =0$) for $s=0$ nm. \cite{BTZ+04, LEL+07} The general
expression for the interaction energy between tubes with axial
magnetization, given by equation (3), can only be solved numerically.
However, tubes that motivated this work \cite{DKG+07, BJK+07, EBJ+08}
satisfy $R/L=\alpha \ll ~1$, in which case one can use that $J_{1}\left(
\alpha x\right) \approx \alpha x/2$. With this approximation, Eq. (3) can be
written in a very simple form as 
\begin{multline}
\tilde{E}_{\mbox{\tiny int}}=-\frac{\sigma _{1}\sigma _{2}R^{2}\left(
1-\beta ^{2}\right) }{4Ld}  \label{eint2} \\
\left[ \frac{1}{\sqrt{1+\left( \frac{L-s}{d}\right) ^{2}}}-\frac{2}{\sqrt{%
1+\left( \frac{s}{d}\right) ^{2}}}+\frac{1}{\sqrt{1+\left( \frac{L+s}{d}%
\right) ^{2}}}\right] \,.
\end{multline}%
The simplicity of Eq.~(\ref{eint2}) makes it an excellent tool for the
understanding of interactions between those nanoelements. Figure 3
illustrates the interaction energy, obtained from Eq. (4), between two
identical nanotubes with parallel axial magnetization as a function of $2R/d$%
. When the two tubes are in contact, $2R/d=1$; when they are infinitely
separated, $2R/d=0$. Differences between this expansion and the full
expression in Eq. (3) are less than $3\%$ for any $L/R$. As an illustration,
when we consider two Ni nanotubes ($\beta =0.8$) with $L=500$ nm, $R=50$ nm, 
$s=100$ nm and $d=100$ nm, we obtain $E{\mbox{\tiny int}}=13.15$ eV from Eq.
(3), and $E{\mbox{\tiny int}}=13.75$ eV from Eq. (4).

\begin{figure}[h]
\begin{center}
\includegraphics[width=8cm]{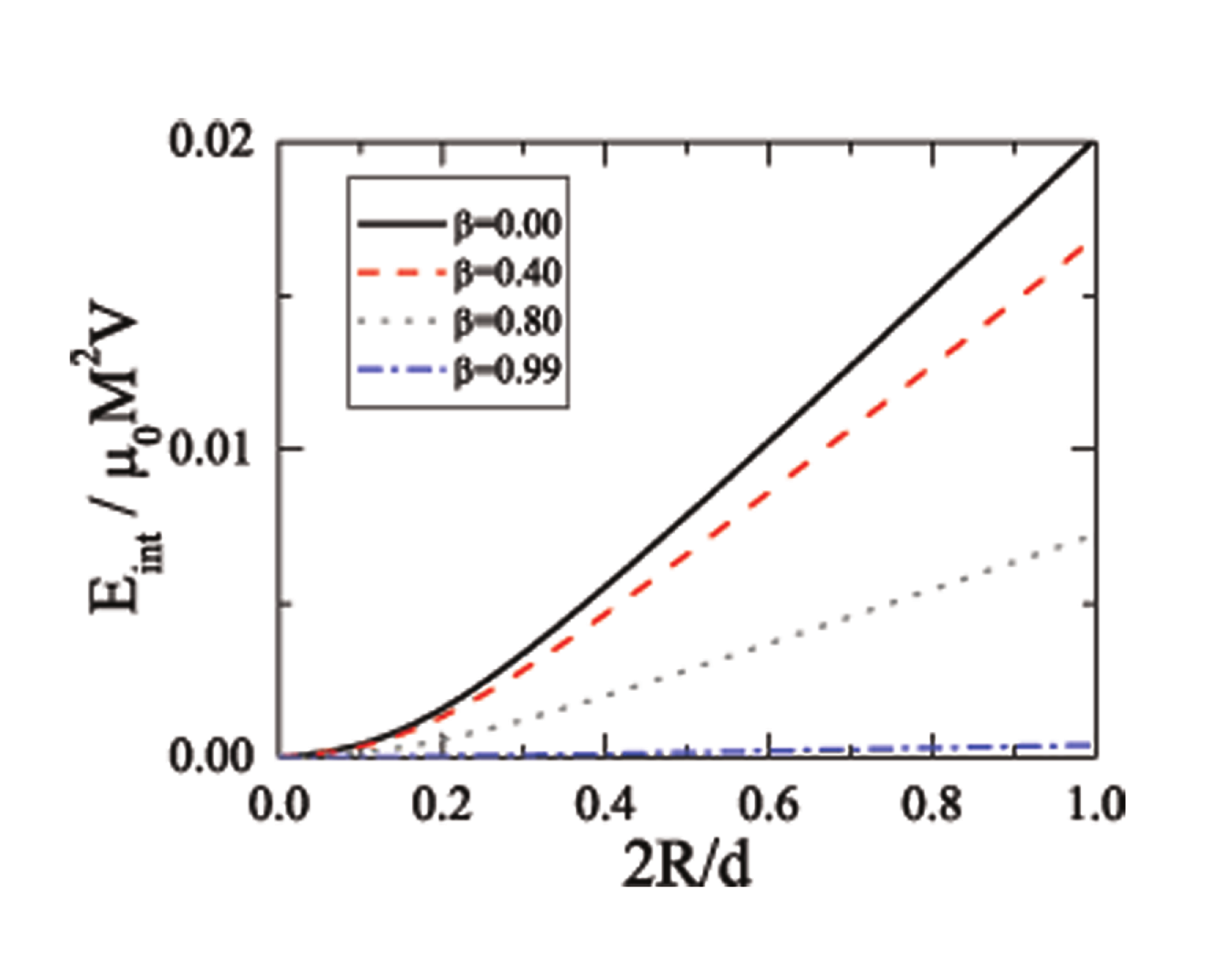}
\end{center}
\caption{(Color online) Interaction energy between two identical nanotubes,
as given by Eq. (4). The tubes have $L=500$ nm, $R=50$ nm, different values
of $\protect\beta$ and parallel magnetization defined by $\protect\sigma %
_{1}=\protect\sigma _{2}$. The vertical position kept fixed, $s=0$ nm, and
the inter-axis distance $d$ is varied as a function of $2R/d$.}
\end{figure}

It is interesting to analyze the behavior of the interaction energy given by
Eq.~(\ref{eint2}) as the inter-axis distance, $d$, is kept fixed and the
vertical separation, $s$, is varied. We define the critical separation $s_{0}
$ such that $\tilde{E}_{\mbox{\tiny int}}(L,d,s_{0})=0$. For tubes with $%
L=500$ nm and $d=100$ nm, $s_{0}=317$ nm, independently of other parameters,
as depicted in Fig. (4). Since lateral positions (small $s$) favor
antiparallel magnetization alignment, the interaction energy is positive for 
$s\le s_{0}=317$ nm. For $s>s_{0}=317$ nm the interaction between tubes with
parallel magnetization is atractive, the strongest attraction appearing for $%
s\approx 500$ nm.

\begin{figure}[h]
\begin{center}
\includegraphics[width=8cm]{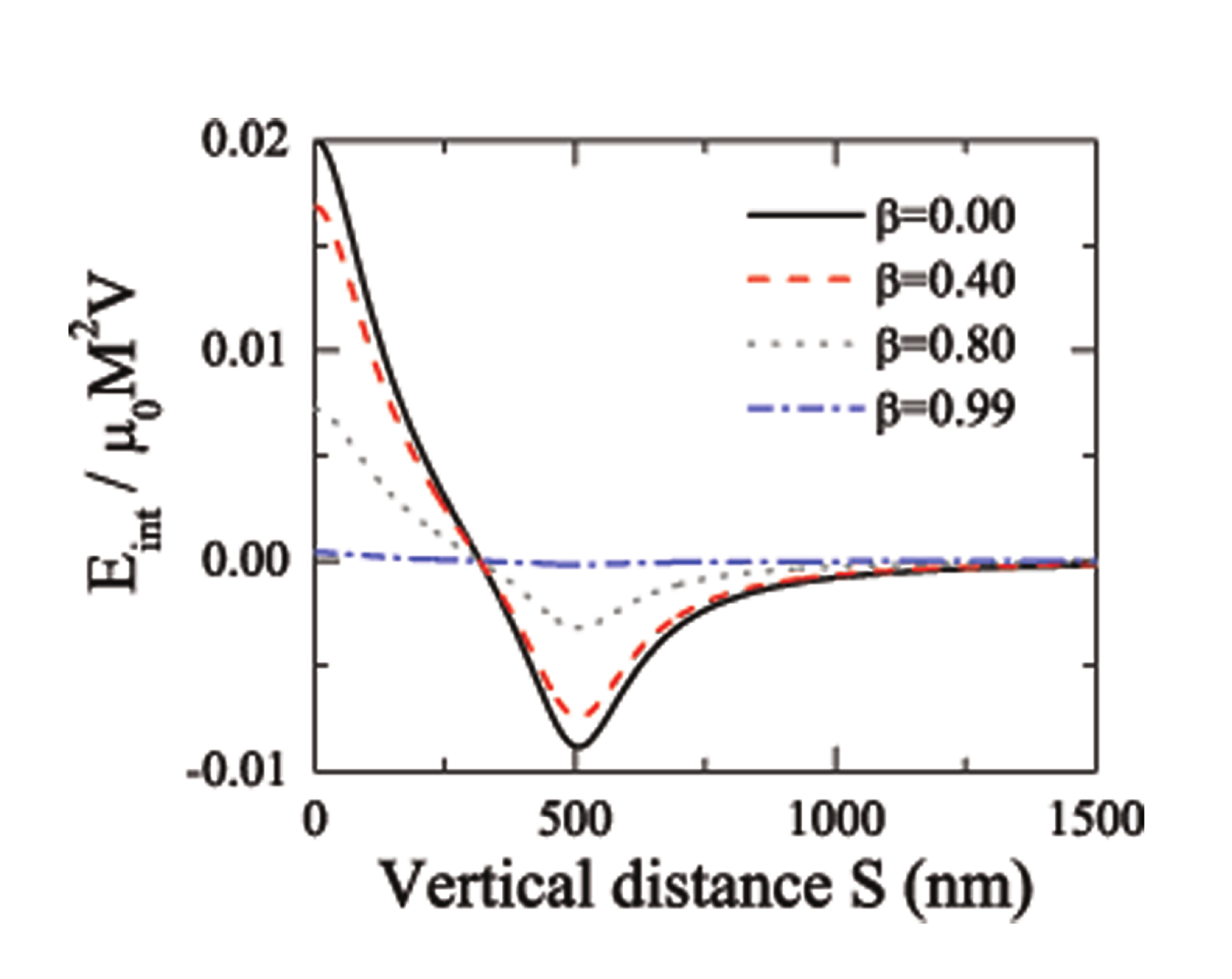}
\end{center}
\caption{(Color online) Interaction energy of two identical nanotubes, as
given by Eq. (4). The tubes have $L=500$ nm, $R=50$ nm, different values of $%
\protect\beta$ and parallel magnetization defined by $\protect\sigma_1=%
\protect\sigma_2$. The inter-axis distance kept fixed, $d=100$ nm, and the
vertical separation $s$ is varied.}
\end{figure}

The dependence of $s_0$ on $L$ for different values of $d$ can be seen in
Fig. (5) For $L\gtrsim 2.5$ $\mu$m, which corresponds to values usually
found in nanotubes, we observe a linear dependence of the form $s_0=0.62L$,
almost independent of $d$.

\begin{figure}[h]
\begin{center}
\includegraphics[width=8cm]{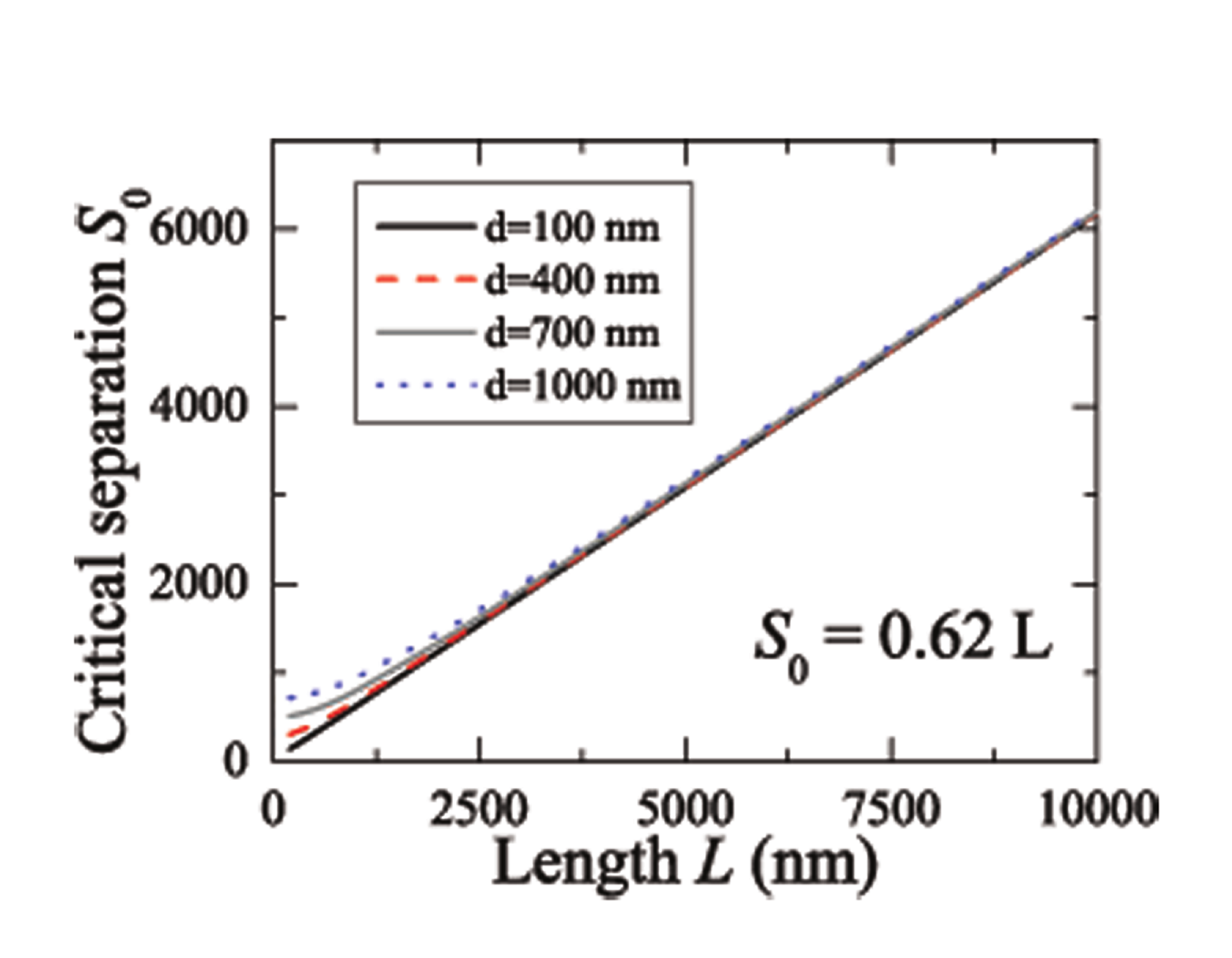}
\end{center}
\caption{(Color online) Dependence of the critical vertical separation $s_0$
on $d$ and $L$ for two identical tubes with arbitrary values of $\protect%
\beta$ and $R$. For $s<s_0$ antiparallel alignment of the magnetization is
favored.}
\end{figure}

In conclusion, by expanding the general expression for the magnetostatic
interaction energy between tubes, and keeping the first-order term we have
obtained an expression that can be easily used to calculate the
magnetostatic interaction between tubes. In particular, we have investigated
the relative position of the tubes for which the ferromagnetic and
antiferromagnetic configurations are of lowest energy. For the tubes usually
found in the literature, $L\gtrsim 2.5$ $\mu$m, we observe a linear
dependence of the critical vertical distance of the form $s_0=0.62L$. Our
results are intended to provide guidelines for the production of
barcode-type nanostructures with prospective applications such as biological
separation and transport.

This work has been partially supported by FONDECYT (N$^\circ$ 11070010 and N$%
^\circ$ 1080300), Millenium Science Nucleus \textit{Basic and Applied
Magnetism} P06-022F, Departamento de Investigaciones Cient\'ificas y
Tecnol\'ogicas, USACH, and AFOSR (award N$^\circ$ FA9550-07-1-0040) in
Chile, and Instituto do Mil\^enio de Nanotecnologia, MCT/CNPq, FAPERJ, and
PROSUL/CNPq in Brazil. CONICYT Ph.D. program and Graduate Direction of
Universidad de Santiago de Chile are also acknowledged.


\begin{thebibliography}{99}
\bibitem{Iijima91} S. Iijima, Nature \textbf{354}, 56-58 (1991).

\bibitem{SCG+04} M. S. Sander, M. J. Cote, W. Gu, B. M. Kile, and C. P.
Tripp, Adv. Mater. \textbf{16}, 2052-2057 (2004).

\bibitem{MLT+02} D. T. Mitchell, S. B. Lee, L. Trofin, N. Li, T. K. Nevanen,
H. Soderlund, and C. Martin, J. Am. Chem. Soc. \textbf{124}, 11864-11856
(2002).

\bibitem{KNS02} A. Kros, R. J. M. Nolte, and N. A. J. M. Sommerdijk, Adv.
Mater. \textbf{14} 1779-1782 (2002).

\bibitem{LMT+02} S. B. Lee, D. T. Mitchell, L. Trofin, T. K. Nevanen, H.
Soderlund, and C. R. Martin, Science \textbf{296}, 2198-2200 (2002).

\bibitem{KHC+04} P. Kohli, C. C. Harrel, Z. Cao, R. Gasparac, W. Tan, and C.
R. Martin, Science \textbf{305}, 984-986 (2004).

\bibitem{SRH+05} S. J. Son, J. Reichel, B. He, M. Schuchman, and S. B. Lee,
J. Am. Chem. Soc. \textbf{127}, 7316-7317 (2005).

\bibitem{DKG+07} M. Daub, M. Knez, U. Goesele, H. Jeske, and K. Nielsch, J.
Appl. Phys. \textbf{101}, 09J111 (2007).

\bibitem{BJK+07} J. Bachmann, J. Jing, M. Knez, S. Barth, H. Shen, S.
Mathur, U. Goesele, and K. Nielsch, J. Am. Chem. Soc. \textbf{129},
9554-9555 (2007).

\bibitem{EBJ+08} J. Escrig, J. Bachmann, J. Jing, M. Daub, D. Altbir, and K.
Nielsch, Phys. Rev. B \textbf{77}, 214421 (2008).

\bibitem{Eisenstein05} M. Eisenstein, Nature Methods \textbf{2}, 484-484
(2005).

\bibitem{XCV+08} J. Xie, L. Chen, V. K. Varadan, J. Yancey, and M.
Srivatsan, Nanotechnology \textbf{19}, 105101 (2008).

\bibitem{LAE+07} P. Landeros, S. Allende, J. Escrig, E. Salcedo, D. Altbir,
and E. E. Vogel, Appl. Phys. Lett. \textbf{90}, 102501 (2007).

\bibitem{LSN+05} W. Lee, R. Scholz, K. Nielsch, and U. Gosele, Angew. Chem.
Int. Ed. \textbf{44}, 6050-6054 (2005).

\bibitem{NFR+01} S. R. Nicewarner-Pe--a, R. G. Freeman, B. D. Reiss, L. He,
D. J. Pe--a, I. D. Walton, R. Cromer, C. D. Keating, and M. J. Natan,
Science \textbf{294}, 137-141 (2001).

\bibitem{HK02} Riccardo Hertel, and Helmut Kronmuller, J. Magn. Magn. Mater. 
\textbf{238}, 185-199 (2002).

\bibitem{Jackson} J. D. Jackson, \textit{Classical Electrodynamics}, 2nd ed.
(Wiley, New York, 1975).

\bibitem{BTZ+04} M. Beleggia, S. Tandon, Y. Zhu, and M. De Graef, J. Magn.
Magn. Mater. \textbf{278}, 270-284 (2004).

\bibitem{LEL+07} D. Laroze, J. Escrig, P. Landeros, D. Altbir, M. Vazquez,
and P. Vargas, Nanotechnology \textbf{18}, 415708 (2007).
\end{thebibliography}
\end{document}